\documentclass[twocolumn]{aastex631}

\begin{document}

\title{Low-Mass Galaxy Interactions Trigger Black Hole Activity}

\correspondingauthor{Marko Mi\'ci\'c}
\email{mmicic@crimson.ua.edu}

\author[0000-0003-3726-5611]{Marko Mi\'ci\'c}
\affiliation{Homer L. Dodge Department of Physics and Astronomy, The University of Oklahoma, Norman, OK 73019, USA}

\author[0000-0003-4307-8521]{Jimmy A. Irwin}
\affiliation{Department of Physics \& Astronomy, University of Alabama, Tuscaloosa, AL 35401, USA}

\author[0000-0001-7069-4026]{Preethi Nair}
\affiliation{Department of Physics \& Astronomy, University of Alabama, Tuscaloosa, AL 35401, USA}

\author[0000-0002-5087-8183]{Brenna N. Wells}
\affiliation{Department of Physics, Duke University, Durham, NC 27708, USA}

\author[0000-0003-0100-0339]{Olivia J. Holmes}
\affiliation{Department of Physics and Astronomy, Purdue University, West Lafayette, IN 47907, USA}

\author[0009-0007-6515-0535]{Jackson T. Eames}
\affiliation{Department of Physics \& Astronomy, University of Alabama, Tuscaloosa, AL 35401, USA}



\begin{abstract}

The existence of high-$z$ over-massive supermassive black holes represents a major conundrum in our understanding of black hole evolution. In this paper, we probe from the observational point of view how early Universe environmental conditions could have acted as an evolutionary mechanism for the accelerated growth of the first black holes. Under the assumption that the early Universe is dominated by dwarf galaxies, we investigate the hypothesis that dwarf-dwarf galaxy interactions trigger black hole accretion. We present the discovery of 82 dwarf-dwarf galaxy pairs and 11 dwarf galaxy groups using the Hubble Space Telescope, doubling existing samples. The dwarf systems span a redshift range of 0.13$<$z$<$1.5, and a stellar mass range of 7.24$<$log(M$_*$/\(M_\odot\))$<$9.73. We performed an X-ray study of a subset of these dwarf systems with Chandra and detected six new AGN, increasing the number of known dwarf-dwarf-merger-related AGN from one to seven. We then compared the frequency of these AGN in grouped/paired dwarfs to that of isolated dwarfs and found a statistically significant enhancement (4$\sigma$-6$\sigma$) in the interacting sample. This study, the first of its kind at the lowest mass scales, implies that the presence of a nearby dwarf neighbor is efficient in triggering black hole accretion. These results open new avenues for indirect studies of the emergence of the first supermassive black holes.

\end{abstract}



\section{Introduction} \label{sec:intro}

Observations of high redshift quasars suggest that supermassive black holes, billions of times more massive than our Sun, were already in place less than 700 million years after the Big Bang \citep{Bogdan2024, Wang_2021, Banados2018, Yang_2020}. The formation and growth paths of these black holes are largely unexplored and remain in the domain of speculation \citep{1994ApJ...432...52L, 2001ApJ...551L..27M, 2003ApJ...596...34B}. One of the mechanisms that can be responsible for triggering the active galactic nuclei (AGN) and subsequent black hole growth is galaxy interactions. 

Various observational, computational, and theoretical works tried to establish the link between galaxy interactions and black hole activity. \cite{2021MNRAS.508.3672P} used hydrodynamical simulations to demonstrate that gravitational and hydrodynamical torques near pericenter passes trigger black hole growth and bursts of black hole accretion that approach Eddington limits. \cite{refId0} studied galaxy mergers up to redshift $z$=0.6 and found that they play a non-negligible to potentially significant role in AGN triggering. \cite{Li_2023} studied the post-merger remnant galaxies and found that they are two to four times more likely to host an AGN when compared to non-interacting galaxies. On the other hand, \cite{Silva_2021} studied mergers of the most massive galaxies and concluded that in the most massive regime, mergers do not play a significant role in AGN activation. However, none of these works probed populations of galaxies that are prevalent in the early Universe. 

Observations of the ultraviolet luminosity function imply that beyond $z$=8 an unseen population of low-mass galaxies overwhelmingly dominates the galaxy population, and as a result dominates the merger rate per unit volume \citep{2015ApJ...813...21M}.  Recent James Webb Space Telescope observations confirmed this hypothesis by discovering that faint dwarf galaxies played a major role in ionizing the Universe during the first billion years \citep{Atek2024}. Even though these high-$z$ low-mass galaxies are beyond our reach, using local dwarf galaxies as high-$z$ analogs and establishing the relationship between dwarf-dwarf interactions and black hole accretion can provide vital clues for understanding how the first black holes evolved and acquired mass.

However, only a small number of dwarf-dwarf galaxy pairs and dwarf galaxy groups have been detected in large-scale surveys \citep{2015ApJ...805....2S, Paudel_2018, 2017NatAs...1E..25S}, and none of them have been reported to have an AGN. Aside from these large-scale surveys, only one confirmed AGN in a dwarf-dwarf merger is known in the whole Universe \citep{Reines_2014, Kimbro_2021}, and two more dual AGN in dwarf-dwarf mergers have been identified as viable candidates \citep{Mićić_2023}. The lack of a comprehensive survey for AGN in dwarf-dwarf mergers makes investigating the early stages of black hole evolution a non-trivial task.

\section{DATA} \label{sec:data}
In this paper, we used the data derived from the 3D-HST survey, a 248-orbit HST Treasury program \citep{2016ApJS..225...27M, 2014ApJS..214...24S}. It covers all five CANDELS fields and provides information on distance (spectroscopic or photometric), stellar masses, star formation rates, and stellar ages for over 200,000 galaxies.  The survey performed the WFC3/G141 grism spectroscopy, yielding precise distance estimates and emission line measurements for $\sim$22,000 galaxies. The accuracy of obtained redshifts was evaluated by comparing them to the available ground-based spectroscopic redshifts. It was found that there is an excellent overall agreement between the two samples. However, the accuracy of individual redshifts might be affected by distance, magnitude, color, or star-formation rate. Stellar masses, and other stellar properties, are derived using the FAST code, assuming the \cite{2003MNRAS.344.1000B} stellar population synthesis model, \cite{2003PASP..115..763C} initial mass function, solar metallicity, and exponentially declining star formation histories. For more detailed information, we refer the reader to \cite{2014ApJS..214...24S}. 

The CANDELS deep fields contain an abundance of existing multiwavelength coverage allowing an in-depth search for AGN. Nevertheless, detecting AGN in dwarf galaxies is notoriously challenging. Since X-ray emission is ubiquitous in AGN, most past works used X-ray observatories to uncover dwarf AGN \citep{Lemons_2015, Pardo_2016, Mezcua_2016, bi10.1093/mnras/staa040}. However, lower-mass galaxies are expected to host lower-mass central black holes that are expected to be dim in X-rays, often requiring hundreds of kiloseconds to detect them even at modest distances \citep{2016A&A...596A..64S}. For that reason, we focus on the GOODS-South field, where the deepest Chandra X-ray observations exist. The GOODS-South field was observed with Chandra 102 times between October 1999 and March 2016, for a total exposure time of 7 megaseconds. The unprecedented depth providing extreme sensitivity coupled with superior angular resolution, makes Chandra an ideal tool to search for dwarf AGN in the GOODS-South field. With the large spread in redshifts in galaxies from the 3D-HST sample, the X-ray detection limits vary greatly. However, throughout the sample, we can probe low luminosity AGN (LLAGN) up to some extent. The detection of LLAGN is important for the completeness of our study because simulations suggest that AGN luminosity is variable on different timescales throughout the merger. The AGN remain in the low-luminosity state (L$_{bol}<$10$^{42}$ erg s$^{-1}$) for $\sim$1-2 gigayears \citep{Hopkins_2008}, significantly longer than in the high-luminosity state (L$_{bol}>$10$^{42}$ erg s$^{-1}$), in which they are expected to spend only $\sim$1-10 megayears \citep{Hopkins_2009}. Therefore, a study unable to probe LLAGN would not achieve an acceptable census of merger-triggered AGN.
\section{RESULTS} \label{sec:res}
\subsection{Dwarf-dwarf pairs and dwarf groups}
We surveyed the publicly available galaxy catalogs from the 3D-HST Survey, consisting of objects residing in GOODS-South, GOODS-North, AEGIS, UDS, and COSMOS deep fields \citep{2016ApJS..225...27M, 2014ApJS..214...24S}. We selected all galaxies with spectroscopically confirmed redshifts and stellar masses less than log(M$_*$/\(\textup{M}_\odot\))$<$9.75 ($\sim$5$\times$10$^9$ \(\textup{M}_\odot\)), resulting in 1663 dwarf galaxies. We then examined the environments of each galaxy, looking for those dwarfs that only had another dwarf companion within $\Delta z$=0.002 and 100 kpc in the projected distance. The $\Delta z$ criterion reflects a very conservative assumption of redshift uncertainty, while the projected separation criterion corresponds to an approximate virial radius of a dwarf galaxy \citep{2015ApJ...805....2S}.  We classify our dwarf systems in five different groups: (1) confirmed dwarf-dwarf pairs, where both dwarfs have spectroscopic redshifts; (2) candidate dwarf-dwarf pairs, where one galaxy has a spectroscopic redshift, while the other has a photometric redshift; (3) confirmed dwarf galaxy groups, where three or more dwarfs have spectroscopic redshifts; (4) confirmed dwarf pairs but potential dwarf groups, where two dwarfs have spectroscopic redshifts and one or more additional galaxies have photometric redshifts; (5) candidate dwarf groups, where one galaxy has a spectroscopic redshift and two or more dwarfs have photometric redshifts. This procedure resulted in the discovery of 82 new dwarf-dwarf galaxy pairs and 11 new dwarf galaxy groups, approximately doubling existing samples.
\subsection{Demographics of dwarf systems}
\begin{figure*}[p]%
\centering
\includegraphics[width=0.9\textwidth]{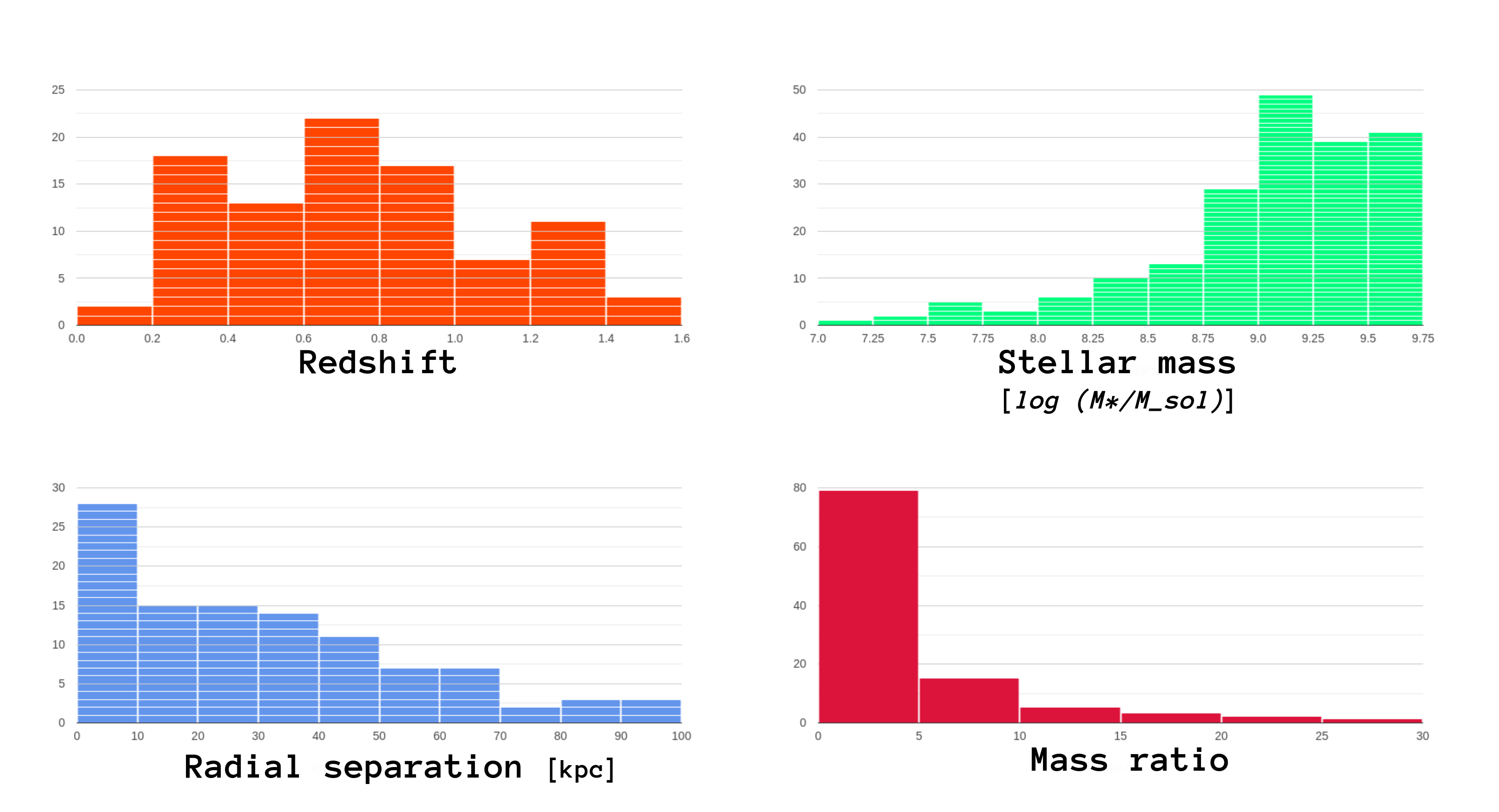}
\caption{Demographics of paired/grouped dwarfs. Upper left: Redshift distribution of paired/grouped dwarf systems. Upper right: Stellar mass distribution of all dwarfs from paired/grouped dwarf systems. Lower left: Distribution of projected separation between the most massive dwarf and their neighbor(s). Lower right: Distribution of the mass ratios of the most massive dwarf and their neighbor(s). }\label{fig1}
\end{figure*}

\begin{figure*}[h]%
\centering
\includegraphics[width=0.9\textwidth]{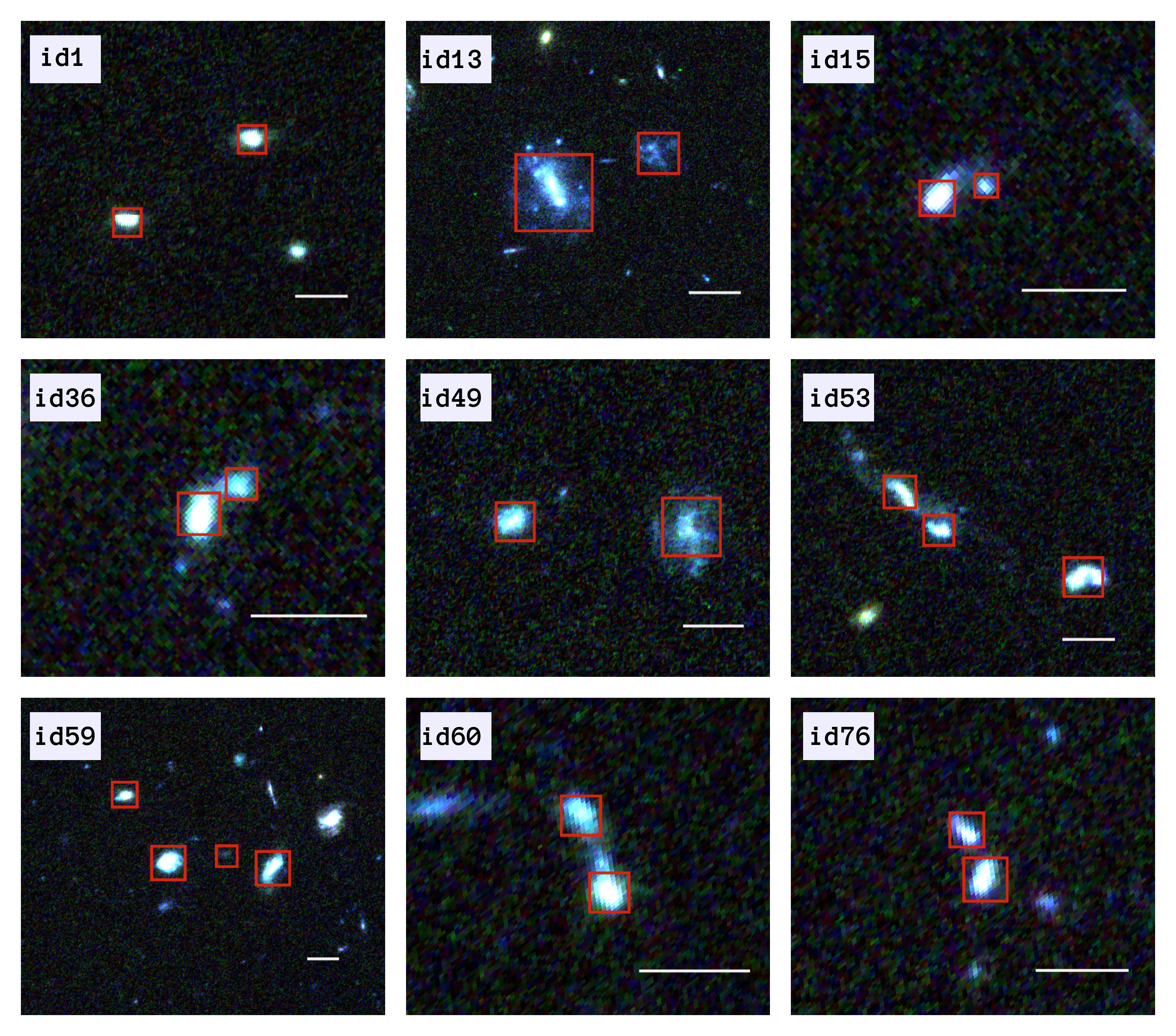}
\caption{GOODS color images of nine randomly selected dwarf systems. The identification number of each system is given in the upper left corner. The white horizontal lines in the lower right corner represent $\approx$10 kpc. The galaxies of interest are labeled with red rectangles.}\label{fig2}
\end{figure*}

The environment of every dwarf galaxy was carefully examined, primarily looking for spectroscopically confirmed dwarf neighbors. However, we also looked for nearby dwarfs with matching photometric redshifts, constituting potential dwarf pairs and groups. This procedure resulted in 51 spectroscopically confirmed dwarf pairs, 31 potential dwarf pairs, 8 spectroscopically confirmed dwarf groups, and 3 potential dwarf groups. Out of 93 dwarf systems, 40 ($\sim$43\%) reside in the GOODS-N field, 29 ($\sim$31.2\%) reside in the GOODS-S field, 17 ($\sim$18.3\%) reside in the AEGIS field, 6 ($\sim$6.5\%) reside in the COSMOS field, and 1 ($\sim$1\%) resides in the UDS field. The field distribution of dwarf systems roughly follows the distribution of the parent dwarf sample. Our dwarf systems span a wide range of redshifts, 0.127$<z<$1.486, projected separations 3 kpc$<r_{sep}<$93 kpc, stellar masses 7.24$<$log(M$_*$/\(M_\odot\))$<$9.73, and mass ratios 1$<M_{1*}/M_{2*}<$26. Such a diverse sample will be of essential importance for future comparative studies of various drivers of different aspects of galactic evolution. The summary of properties of all dwarf systems is given in Tables \ref{tab1} through \ref{tab3}. The demographics of dwarf systems is graphically represented in Figure \ref{fig1}. A selected subsample of dwarf systems is shown in Figure \ref{fig2}.

\subsection{X-ray analysis}
We crossmatched all dwarf galaxies from the GOODS-South dwarf systems with the Chandra X-ray Catalog \citep{2010ApJS..189...37E} and found seven X-ray sources within 1$\arcsec$  from the host dwarf. The X-ray sources reside in dwarf systems labeled as id3, id10, id34, id43, id81, id86, and id91 in Table \ref{tab1} through Table \ref{tab3}. Observed fluxes in the 0.3-7 keV band vary from 1$\times$10$^{-14}$ to 3$\times$10$^{-17}$ erg s$^{-1}$ cm$^{-2}$. The X-ray luminosities were calculated using reported fluxes and adopting the redshift of the host galaxy as distance:
\begin{equation}
    L_X=4\pi d_L^2f_X
\end{equation}
where d$_\textup{L}$ is the luminosity distance in centimeters and f$_\textup{X}$ is X-ray flux, and then converted to rest-frame 0.5-10 keV luminosities assuming a model with a typical AGN power law of $\Gamma$=1.9 and accounting for the appropriate K-correction factor. The luminosities range from 2$\times$10$^{40}$ to 3$\times$10$^{43}$ erg s$^{-1}$. We note that six out of seven X-ray sources have luminosities $<$10$^{42}$ erg s$^{-1}$, in line with LLAGN. Such low luminosities do not necessarily have to be produced by an accreting supermassive black hole. For example, the hot interstellar medium (ISM) gas and unresolved populations of X-ray binaries (XRBs) can produce LLAGN-similar levels of X-ray emission. Thus, we evaluate the non-AGN contributions to the observed X-ray flux.

For the X-ray binary contribution, we use the relation from \cite{Lehmer_2016} that is based on a sample of normal galaxies at redshifts $z\sim$0-7:
\begin{equation}
    L^{\textup{XRB}}_{\textup{2-10keV}}=\alpha_O(1+z)^{\gamma}M_*+\beta_0(1+z)^{\delta}SFR
\end{equation}
in erg s$^{-1}$, where log $\alpha_0$=29.30$\pm$0.28, log $\beta_{0}$=39.40$\pm$0.08, $\gamma$=2.19$\pm$0.99, $\delta$=1.02$\pm$0.22, M$_*$ is stellar mass, and SFR is star formation rate in units of \(\textup{M}_\odot\) yr$^{-1}$. The  M$_*$ term is proportional to low-mass XRBs, while the SFR term is proportional to high-mass XRBs. These luminosities are then converted to the 0.5-10 keV energy band assuming a $\Gamma$=1.4 model \citep{Hickox_2006} and applying appropriate K-correction factor. We find that one X-ray source, id81, is in line with the expected emission from the XRB population. On the other hand, the X-ray source from id43, the only non-LLAGN source, exceeds the expected XRB contribution by a factor of $\sim$500. In the remaining five sources, the XRB contributions account for 4\%-22\% of the observed X-ray emission, with a median value of 16\%, requiring an additional source of intensive X-ray emission. However, this relation was derived considering all galaxies with L$_X<$ 3$\times$10$^{42}$ erg s$^{-1}$ to be non-AGN star-forming galaxies. An overwhelming majority of known X-ray-detected dwarf AGN, including six out of seven of our sources, have luminosities lower than this value and would be classified as non-AGN galaxies in the computation of \cite{Lehmer_2016} relation.

Another scaling relation between XRB emission, M$_*$, and SFR is derived by \cite{Lehmer_2010} using a sample of luminous star-forming galaxies:
\begin{equation}
    L^{\textup{XRB}}_{\textup{2-10keV}}=(9.05\pm0.37)\times10^{28}M_*+(1.62\pm0.22)\times10^{39}SFR
\end{equation}
Even though this relation yields lower XRB contributions, the overall results remain the same. One X-ray source, id81, is consistent with the XRB emission; the only non-LLAGN source, id43, exceeds the XRB predicted emission by a factor of 1500; and in the other five sources, the XRB contributions are expected to account for 0.6\%-12\% with a median value of 5\%. We note that various other relations for estimating the XRB emission exist \citep[e.g.][]{Fragos_2013, Lehmer_2022} which may yield different values for the XRB contributions, but the overall conclusions remain the same.

We estimate the 0.5-2 keV hot ISM gas X-ray luminosity using the relation from \cite{10.1111/j.1365-2966.2012.21831.x}:
\begin{equation}
    L^{\textup{hot}}_{\textup{0.5-2keV}}=(8.3\pm0.01)\times10^{38}SFR
\end{equation}
The resulting luminosities are converted to a 0.5-10 keV energy band assuming a power-law index $\Gamma$=3 and using an appropriate K-correction factor. We find that the ISM X-ray emission plays a minor role as it accounts for a fraction of a percent up to a few percent of observed X-ray emission.

\begin{figure*}[p]%
\centering
\includegraphics[width=0.9\textwidth]{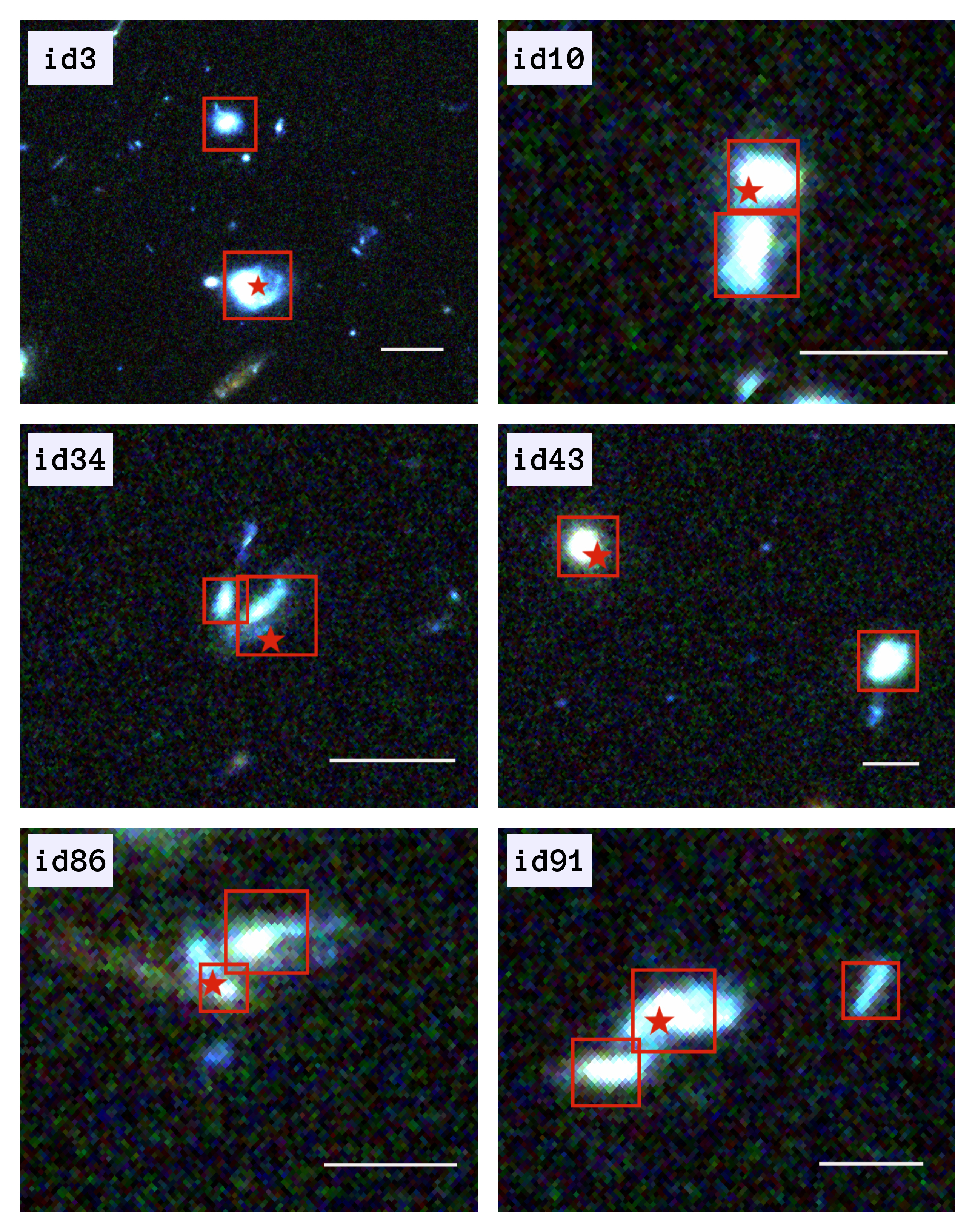}
\caption{GOODS color images of the six dwarf pairs/groups hosting an AGN. The identification number of each system is given in the upper left corner. Horizontal white lines in the lower right corner represent $\approx$10 kpc. Galaxies of interest are located in red rectangles. Red star symbols represent the location of Chandra X-ray detected AGN.}\label{fig3}
\end{figure*}

Therefore, we find that in one case, id81, the expected XRB contribution to the X-ray emission exceeds the observed X-ray emission, thus, the presence of an AGN is unlikely and id81 is removed from the final AGN sample. In the remaining six sources we find sufficient evidence for AGN presence. The total X-ray luminosities, XRB and ISM contributions, and calibrated AGN luminosity for each of the six sources are given in Table \ref{tab4}. Dwarf systems hosting the AGN are shown in Figure \ref{fig3}.
\subsection{AGN frequency}
We note that four out of six of our AGN reside in late-stage galaxy mergers, where galaxies are separated by less than 10 kpc. Due to the large redshifts, the angular separations correspond to $<$1\arcsec. This implies that even if the observations were perfectly focused, distinguishing between the two AGN would be a non-trivial task. We find that the smallest off-axis angle is $\sim$2\farcm5 (id86), and the resulting Chandra point-spread function is greater than the angular separation between the merging galaxies, leaving the possibility of a dual AGN appearing as a single AGN. Various works agree that the dual AGN frequency increases with the decrease of galaxy separation \citep[e.g.,][]{Koss_2012,Barrows_2017}, implying that probably some, if not all, of our four late-stage mergers host dual AGN. As a result of this ambiguity, we report two AGN values: the lower AGN limit consisting of six confirmed AGN, N$_{\textup{agn,low}}$=6; and the higher AGN limit, assuming that all of the four late-stage mergers host dual AGN, N$_{\textup{agn,high}}$=10. The corresponding X-ray AGN frequency per paired/grouped dwarf galaxy is 9.8\%$^{+5.9}_{-3.9}$ in the lower limit, and 16.4\%$^{+7.0}_{-5.1}$ in the higher limit, with uncertainties obtained from the Poissonian statistics at 1$\sigma$ level. \citep{1986ApJ...303..336G}. In follow-up work, we plan to use novel Bayesian tools such as BAYMAX to estimate the likelihood of dual AGN posing as single AGN \citep{Foord_2019}.
\subsection{Control sample}
The inferred AGN frequency in interacting dwarfs, in either limit, is significantly higher than in any existing work that tackled the problem of dwarf AGN. However, many factors can play a role in different AGN frequencies, the most prominent reasons being redshifts, stellar masses of galaxies examined, and the depth of observations used. To truthfully estimate the magnitude of enhanced black hole accretion (if any) in an interacting sample of dwarf galaxies, we constructed a sample of isolated dwarfs with identical redshift and stellar mass distributions, for which the same depth of X-ray observations are available. Therefore, the interacting dwarfs and control sample dwarfs would be intrinsically more-or-less identical, exposed to the same observational conditions, and the only variable that could cause different AGN frequency would be the environment, i.e., the relative proximity of another dwarf galaxy.

\begin{figure*}[p]%
\centering
\includegraphics[width=0.5\textwidth]{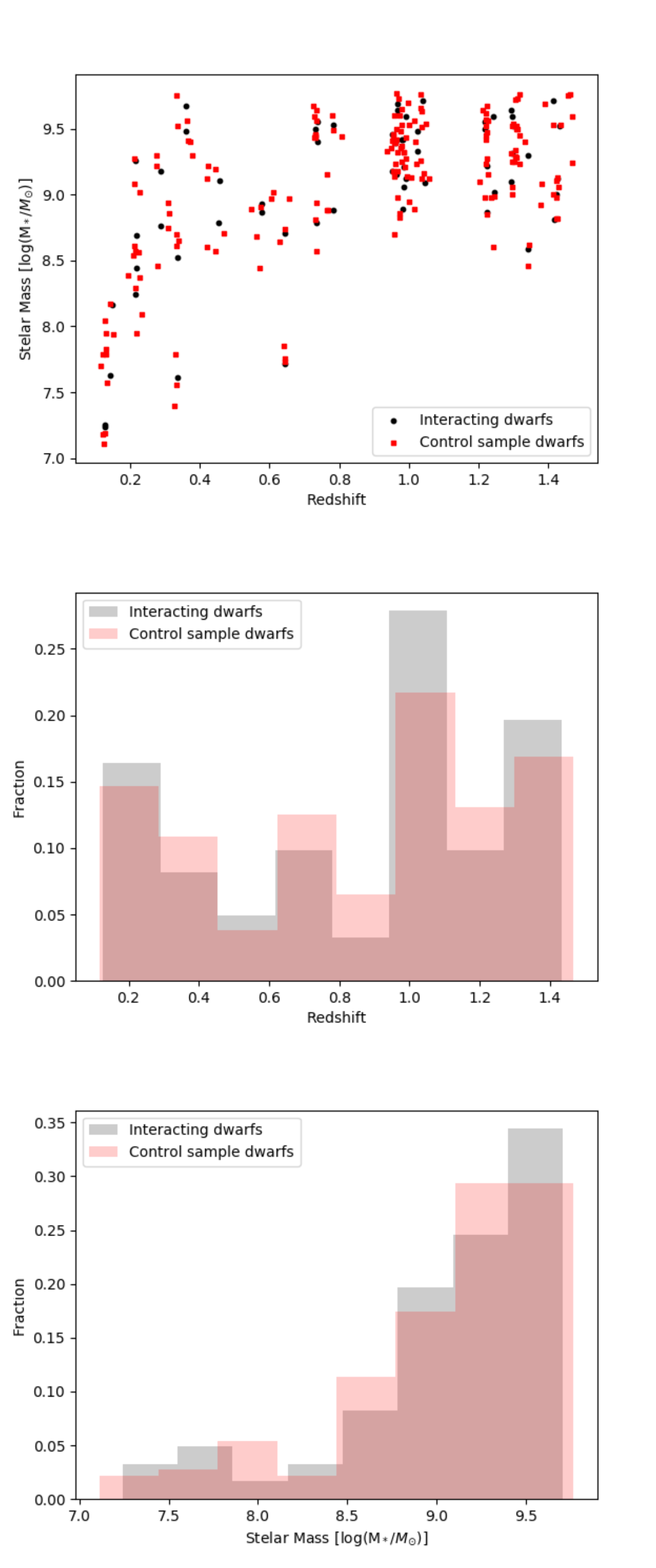}
\caption{Top: Distribution of interacting and isolated control dwarf galaxies in redshift-stellar mass space. Middle: Redshift distribution of interacting and isolated control dwarf galaxies. Bottom: Stellar mass distribution of interacting and isolated control dwarf galaxies. }\label{fig4}
\end{figure*}

Specifically, for each interacting dwarf galaxy, we selected three dwarf galaxies within $\Delta z$=0.05 and $\Delta$log(M$_*$/\(M_\odot\))=0.3, residing in the GOODS-South field, for which the same depth of X-ray observations can be achieved, without spectroscopically confirmed neighbors within 200 kpc. For each potential control sample dwarf, we also examined available archival imaging and excluded those dwarfs that are involved in an ongoing merger with uncataloged neighbors or dwarfs exhibiting clear tidal debris that might be due to ongoing or previous interaction \citep{2020AJ....159..103K}. We started by selecting spectroscopically confirmed isolated dwarfs most similar to their interacting counterparts in terms of redshift and stellar mass. However, in some cases due to the combination of high-$z$ and low stellar mass, we were unable to locate appropriate spectroscopically confirmed isolated dwarfs (e.g., lower-mass companions in id36, id70). In these cases, we included dwarfs with well-constrained photometric redshifts, i.e., those dwarfs with redshift uncertainties estimated at the 95\% confidence limits meeting the criterion of $|z_{upper/lower}-z_{phot}|<$ 0.05. These modest variations of redshift do not drastically affect the stellar mass estimates nor the presumed isolation of control sample dwarfs. We find, in total, 183 unique isolated control sample dwarfs, three for each of the 61 interacting dwarfs. A comparison between interacting and isolated control sample dwarfs is shown in Figure \ref{fig4}, where it can be seen that they follow the same distribution in the stellar mass-redshift space.

We then performed an identical search for AGN in the isolated control sample. We crossmatched the control sample dwarfs with the Chandra Source Catalog, looking for X-ray sources within 1$\arcsec$ of each dwarf, and we found three matches implying AGN frequency of 1.6\%$^{+1.6}_{-0.8}$. Various other works examined the AGN occupation fractions in dwarf galaxies using multiwavelength approaches. For example, \cite{2013ApJ...775..116R} found a 0.5\% occupation fraction at optical wavelengths. This value likely underestimates the true occupation fraction because of the combination of selection biases and limitations of existing optical diagnostics methods. The sample of dwarf galaxies was derived from a shallow SDSS survey, which, except for the nearest dwarfs, favors more luminous dwarfs. The enhanced luminosity usually signals a prevalent younger stellar population and ongoing star formation. Active star formation tends to overpower the optical signatures of AGN, thus rendering the optical BPT diagnostics methods \citep{1981PASP...93....5B} inaccurate for these types of works. On the other hand, \cite{k10.1093/mnrasl/slz102} used infrared observations to find that AGN occupation fraction in dwarf galaxies can vary from 10\% to 30\%. \cite{l10.1093/mnras/stz3636} reanalyzed this work, taking into account the poor angular resolution of the infrared observations and the well-known fact that star formation processes can mimic typical AGN signatures in dwarf galaxies \citep{Hainline_2016}, and found an occupation fraction of only 0.4\%. Some recent works suggest that dwarf galaxies indeed have significantly higher AGN fractions. \cite{mezc10.1093/mnras/stae292} studied dwarf galaxies from the MaNGA survey using a spaxel-by-spaxel classification in three spatially-resolved emission line diagnostic diagrams. This approach resulted in an AGN fraction of 20\%, indicating that integral field spectroscopy is a powerful tool to reveal previously unknown faint and low-accreting dwarf AGN. \cite{Dickey_2019} performed long-slit spectroscopy of 20 isolated, quenched dwarf galaxies, where star-formation pollution is expected to be minimal, and found a remarkable 80\% AGN fraction. \cite{Pacucci_2021} developed a theoretical model that utilizes the physical properties of galaxies, such as stellar mass and angular momentum, to predict multiwavelength AGN fractions of 5\%-30\%, increasing with increased stellar mass and decreased metallicity.

The apparent ambiguity of dwarf AGN fractions in existing literature is not unexpected. The use of different wavelengths and approaches will achieve different AGN luminosity depths and thus probe different AGN populations. In order to properly evaluate the validity of our control sample AGN fraction, we must focus on X-ray studies probing comparable AGN luminosity levels. \cite{Latimer_2021} used the eROSITA telescope, at X-ray wavelengths, to find an upper limit on dwarf galaxy AGN incidence of 1.8\%. \cite{2018MNRAS.478.2576M} studied dwarfs from the Chandra COSMOS Legacy Survey and found occupation fraction of $\sim$0.4\% at $z<$0.4 and L$_X>$10$^{41}$ erg s$^{-1}$. Similarly, \cite{Pardo_2016} used Chandra to find dwarf AGN fractions of 0.6\%-3\% at $z<$0.6 and L$_X>$10$^{41}$ erg s$^{-1}$. Finally, \cite{bi10.1093/mnras/staa040} studied SDSS dwarf galaxies at X-ray wavelengths with XMM-Newton and found AGN incidence to be between $\sim0.5\%$ in the least massive stellar-mass bin, up to $\sim1\%$ in the most massive stellar-mass bin calculated at L$_X>$3$\times$10$^{40}$. When a lower limit of L$_X>$10$^{39}$ is considered, the AGN fraction increases from 2.5\% to 5.5\% in the lowest and highest mass bins, respectively. If detection limits are further reduced, the AGN fraction would likely grow. However, reaching extremely low X-ray luminosity detection limits for a large sample of dwarfs would be observationally expensive. Furthermore, such very low-luminosity AGN would be almost impossible to distinguish from stellar-mass X-ray emitters with current technical capabilities. This review of the literature suggests that even though certain approaches yield large dwarf AGN fractions due to their superior sensitivity, the X-ray-based works consistently find low AGN fractions, in accordance with the AGN fraction we found in our control sample.

\begin{figure*}[p]%
\centering
\includegraphics[width=1\textwidth]{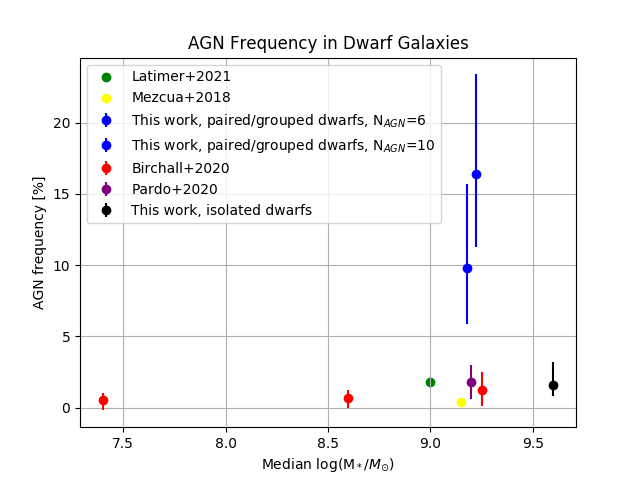}
\caption{AGN frequency in paired/grouped dwarfs in both N$_{AGN,low}$=6 and N$_{AGN,high}$=10 limits compared to the occupation fraction in isolated control sample and occupation fraction from the literature as a function of the median stellar mass of dwarfs hosting an AGN. Uncertainties from our work are estimated at 1$\sigma$ assuming Poissonian statistics. \cite{Latimer_2021} does not provide the list of galaxies in their sample, thus the median stellar mass of the sample is set to log (M$_*$/\(\textup{M}_\odot\))=9.0}\label{fig5}
\end{figure*}

Therefore, the AGN frequency we found in the interacting sample indicates enhanced AGN presence by a factor of six up to a factor of ten, depending on the limit observed. The statistical significance of our discovery is from $\sim$4.3$\sigma$ in the lower limit of the number of interacting AGN, up to $\sim$6.7$\sigma$ in the AGN upper limit scenario. We conclude that dwarf-dwarf galaxy interactions play a significant role in triggering activity and subsequent growth of black holes. A comparison between AGN frequency we found in interacting dwarf galaxies and isolated control sample dwarf galaxies, and various X-ray-based works from the literature is shown in Figure \ref{fig5}.
\section{Conclusions}
The idea that galaxy interactions trigger AGN in dwarfs is not novel. Different simulations studied minor mergers involving galaxies with disproportional masses where one galaxy could be a dwarf. It was found that the secondary AGN can react dramatically to the interaction and undergo brief but violent episodes of accretion that can increase the mass of the black hole by ten times, representing a viable channel for rapid black hole growth \citep[][and references therein]{Pfeifle_2023}. Some recent observational works managed to detect mergers of dwarfs and their more massive companions with secondary AGN dominating the primary, suggesting ongoing intensive accretion processes \citep{Secrest_2017, Mi_i__2024}. However, in the cosmological context, dwarf-dwarf mergers play an integral role in understanding the early evolutionary paths of supermassive black holes. Even though conditions in the early Universe remain beyond our reach, the prevailing wisdom is that concurrent and consecutive mergers involving dwarf galaxies are a common sight. In this paper, we presented a new sample of dwarf galaxy systems, the first sample of dwarf-dwarf-merger-related AGN, and unambiguous evidence that dwarf-dwarf mergers are efficient in triggering black hole accretion. These findings are, for the first time, opening a window for future studies of mechanisms that dictate the earliest stages of growth of supermassive black holes.
\begin{acknowledgments}
The authors thank the anonymous referee for providing insightful comments that greatly improved the manuscript.
\end{acknowledgments}
\startlongtable
\begin{deluxetable*}{ccccccccc}
\tablenum{1}
\tablecaption{Summary of spectroscopically confirmed dwarf-dwarf pairs\label{tab1}}
\tablewidth{0pt}
    \tablehead{\colhead{ID} & \colhead{RA}  & \colhead{Dec} & \colhead{Field}&\colhead{$z$} & \colhead{$z$ source} &\colhead{log M$_*$}&\colhead{r$_{\textup{sep}}$}&\colhead{M$_{1*}$/M$_{2*}$} \\
    \colhead{} & \colhead{(deg)}  & \colhead{(deg)} & \colhead{}&\colhead{} & \colhead{} &\colhead{(\(\textup{M}_\odot\))}&\colhead{(kpc)}&\colhead{}  \\ \hline
    \colhead{(1)} & \colhead{(2)}  & \colhead{(3)} & \colhead{(4)}&\colhead{(5)} & \colhead{(6)} &\colhead{(7)}&\colhead{(8)}&\colhead{(9)}
}
\startdata
id1    & 189.04290771&	62.19090271  & GOODS-N&0.7728&s&9.41& &\\ 
    & 189.04084778&	62.19153976& GOODS-N  & 0.7732&s&9.14 &31&1.9:1 \\ \hline
id2   & 53.2062912&	--27.85160446  & GOODS-S&0.7836&s&9.53& & \\
   &53.20492554&	--27.85132599  & GOODS-S&0.7838&s&8.88&34 &4.5:1  \\ \hline
id3   & 53.07991409	&--27.74620056   & GOODS-S&0.2180&s&8.69&  &\\
   & 53.08021927&	--27.74412537   & GOODS-S&0.2170&s&8.44&27&1.8:1 \\ \hline
id4   & 189.37626648&	62.32131577   & GOODS-N&0.8357&s&9.34& & \\
   & 189.37515259	&62.32249832   & GOODS-N&0.8360&s&9.15&36&1.5:1 \\ \hline
id5   & 189.151474&	62.27391815   & GOODS-N&0.8520&s&9.68& & \\
   & 189.15444946&	62.27086258   & GOODS-N&0.8518&s&9.36&94&2.1:1 \\ \hline
id6   & 189.16722107&	62.2768631   & GOODS-N&0.8490&s&9.66& & \\ 
   & 189.1685791&	62.27826691   & GOODS-N&0.8486&s&9.00&43&4.6:1 \\ \hline
id7   & 53.09305954	&--27.89606667   & GOODS-S&0.9666&s&9.69& & \\
   & 53.09386063	&--27.8982563   & GOODS-S&0.9670&s&9.64&67&1.1:1 \\ \hline
id8   & 53.21165085	&--27.89239502   & GOODS-S&0.9860&s&9.21& & \\
   & 53.21179199	&--27.89239502   & GOODS-S&0.9860&s&9.06&4&1.4:1 \\ \hline
id9   & 189.2885437	& 62.33514786   & GOODS-N&1.0111&s&9.71& & \\
   & 189.29125977&	62.33637619   & GOODS-N&1.0107&s&9.00&52 &5.1:1\\ \hline
id10   & 53.06583786&	--27.77513313   & GOODS-S&1.0230&s&9.48&  &\\
   & 53.0657959	&--27.77492905   & GOODS-S&1.0230&s&9.33&6&1.4:1 \\ \hline
id11   & 189.115448&	62.16436386   & GOODS-N&1.2146&s&9.57&&  \\
   & 189.11901855	&62.16487122   & GOODS-N&1.2160&s&9.02&53&3.5:1 \\ \hline
id12   & 53.06857681	&--27.78397751 & GOODS-S&1.2187&s&9.55&&  \\
   & 53.06773376	&--27.78424454   & GOODS-S&1.2190&s&9.50&24&1.1:1 \\ \hline
id13   & 189.11444092&	62.17165375 & GOODS-N&0.2532&s&9.07&&  \\
   & 189.11122131	&62.17215729   & GOODS-N&0.2540&s&7.92&23&14.1:1 \\ \hline
id14   & 189.1065979	&62.14515686 & GOODS-N&1.2630&s&9.46& & \\
   & 189.10676575&	62.14500046   & GOODS-N&1.2630&s&9.34&5&1.3:1 \\ \hline
id15   & 53.15553665&	--27.91340637 & GOODS-S&1.3410&s&9.30& & \\
   & 53.15535736&	--27.91337013   & GOODS-S&1.3410&s&8.59&5&5.1:1 \\ \hline
id16   & 53.20502853	&--27.74331856   & GOODS-S&0.2164&s&9.26& & \\
   & 53.20407486	&--27.74143219 & GOODS-S&0.2162&s&8.24&26&10.5:1 \\ \hline
id17   & 214.79377747&	52.81565857 & AEGIS&0.2837&s&9.39& & \\
   & 214.78594971&	52.81204605   & AEGIS&0.2837&s&8.38&93&10.2:1 \\ \hline
   id18   & 189.17329407	&62.15087128 & GOODS-N&0.2754&s&8.51& & \\
   & 189.16664124	&62.14615631   & GOODS-N&0.2758&s&8.08&88&2.7:1 \\ \hline
id19   & 189.25950623&	62.17575836 & GOODS-N&0.2762&s&7.72& & \\
   & 189.25942993	&62.17593384   & GOODS-N&0.2760&s&7.40&3&2.1:1 \\ \hline
id20   & 150.15498352&	2.32654262 & COSMOS&0.2831&s&9.53&&  \\
   & 150.15553284&	2.32628226   & COSMOS&0.2832&s&9.07&9&2.9:1 \\ \hline
id21   & 53.0802536&	--27.80839348 & GOODS-S&0.2880&s&9.18&&  \\
   & 53.07812881	&--27.8094635   &GOODS-S&0.2870&s&8.76&34&2.6:1 \\ \hline
id22   & 214.73500061	&52.73179245 & AEGIS&0.2909&s&9.43&&  \\
   & 214.73638916&	52.73101807   &AEGIS&0.2909&s&9.41&18&1:1 \\ \hline
id23   & 214.82722473&	52.82629013 & AEGIS&0.2962&s&8.09&&  \\
   & 214.83213806&	52.82337189   &AEGIS&0.2962&s&8.08&67&1:1 \\ \hline
id24   & 189.24482727&	62.21456146 & GOODS-N&0.3201&s&9.40&&  \\
   & 189.2419281	&62.21679306  &GOODS-N&0.3204&s&8.41&44&9.8:1 \\ \hline
id25   & 53.22335434&	--27.8624115 & GOODS-S&0.3352&s&8.52& & \\
   &53.22351837&	--27.86248207   &GOODS-S&0.3352&s&7.61&3&8.1:1 \\ \hline
id26   & 214.95265198&	52.93842697 & AEGIS&0.3134&s&9.67& & \\
   &214.95510864&	52.93862915   &AEGIS&0.3122&s&8.35&25&20.9:1 \\ \hline
id27   &214.72950745	&52.7573967 & AEGIS&0.4043&s&9.49&&  \\
   &214.72917175	&52.75977707   &AEGIS&0.4042&s&8.77&47&5.2:1 \\ \hline
id28   &189.12519836	&62.20683289 & GOODS-N&0.4106&s&8.75& & \\
   &189.12475586&	62.20670319  &GOODS-N&0.4100&s&7.70&5&11.2:1 \\ \hline
id29   &53.16929245&	--27.93000412 & GOODS-S&0.1274&s&7.45&&  \\
   &53.16825104&	--27.93003845 &GOODS-S&0.1270&s&7.24&8&1.6:1 \\ \hline
id30   &189.24327087&	62.26353455 & GOODS-N&0.4565&s&9.30& & \\
   &189.24397278&	62.26481628 &GOODS-N&0.4567&s&8.37&28&8.5:1 \\ \hline
id31   &189.38896179&	62.23109055 & GOODS-N&0.5105&s&9.50& & \\
   &189.38467407&	62.22917175 &GOODS-N&0.5101&s&8.66&62&6.9:1 \\ \hline
id32   &189.21343994&	62.18349075& GOODS-N&0.5174&s&9.00& & \\
   &189.21533203&	62.18114853 &GOODS-N&0.5180&s&8.78&56&1.7:1 \\ \hline
id33   &34.37569809	&--5.22611809& UDS&0.5340&s&9.36& & \\
   &34.37714386	&--5.22825003 &UDS&0.5340&s&9.27&59&1.2:1 \\ \hline
id34   &53.04432297&	--27.78720093& GOODS-S&0.5781&s&8.93& & \\
   &53.04456711	&--27.78714943 &GOODS-S&0.5781&s&8.87&5&1.1:1 \\ \hline
id35   &189.26983643&	62.27832413& GOODS-N&0.6041&s&9.03&&  \\
   &189.2701416&62.27870178 &GOODS-N&0.6041&s&8.83&10&1.6:1 \\ \hline
   id36   &53.07033539&	--27.71787071& GOODS-S&0.6442&s&8.71&&  \\
   &53.07018661	&--27.7177639&GOODS-S&0.6442&s&7.72&4&9.8:1\\ \hline
id37   &150.08889771&	2.3513546& COSMOS&0.6684&s&9.05&&  \\
   &150.08886719	&2.35152078&COSMOS&0.6684&s&8.51&4&3.5:1 \\ \hline
id38   &189.02261353&	62.21254349& GOODS-N&0.6789&s&8.89& & \\
   &189.02279663	&62.21220016&GOODS-N&0.6789&s&8.39&9&3.2:1 \\ \hline
id39   &189.35325623	&62.26233292& GOODS-N&0.6912&s&9.61& & \\
   &189.35220337&	62.26305389&GOODS-N&0.6912&s&9.58&23&1.1:1 \\ \hline
id40   &214.97157288&	52.93635559& AEGIS&0.7164&s&9.58&&  \\
   &214.96983337	&52.93461227&AEGIS&0.7162&s&8.92&53&4.6:1 \\ \hline
id41   &215.00022888&	52.96072006& AEGIS&0.7439&s&9.06& & \\
   &214.99899292&	52.96024323&AEGIS&0.7437&s&8.98&24&1.2:1 \\ \hline
id42   &214.76210022	&52.77083588& AEGIS&0.7380&s&9.20&&  \\
   &214.76611328&	52.7721405&AEGIS&0.7379&s&8.85&73 &2.2:1\\ \hline
id43   &53.12589645	&--27.75127602& GOODS-S&0.7380&s&9.55& & \\
   &53.12376785	&--27.75199509&GOODS-S&0.7370&s&9.40&54&1.4:1 \\ \hline
id44   &214.78341675&	52.84712601& AEGIS&0.7567&s&9.72&&  \\
   &214.78218079	&52.84781647&AEGIS&0.7568&s&8.44&27&19.1:1 \\ \hline
id45   &189.26382446&	62.27958679& GOODS-N&0.7439&s&9.32&&  \\
   &189.26036072&	62.2792244&GOODS-N&0.7441&s&8.79&44&3.4:1 \\ \hline
id46   & 189.28974915&	62.19259644 & GOODS-N&0.9064&s&9.73& & \\
   &189.28616333&	62.19122696   &GOODS-N&0.9072&s&9.48&62 &1.8:1\\ \hline
id47   &189.03062439	&62.16661072
 & GOODS-N&0.9229&s&9.16&&  \\
   &189.0272522	&62.16641998   &GOODS-N&0.9223&s&8.91&45&1.8:1 \\ \hline
id48   &53.14984894	&--27.85508728& GOODS-S&0.3592&s&9.67& & \\
   &53.15156555&	--27.8548851  &GOODS-S&0.3588&s&9.48&28&1.5:1 \\ \hline
id49   &189.12135315	&62.19810867 & GOODS-N&0.5293&s&9.00& & \\
   &189.12417603&	62.19818878 &GOODS-N&0.5285&s&8.62&30&2.4:1 \\ \hline
id50   &214.89303589&	52.7860527
 & AEGIS&0.6518&s&9.49&&  \\
   &214.89216614	&52.78372574 &AEGIS&0.6528&s&9.17&60&2.1:1 \\ \hline
   id51   &150.14634705	&2.26014423& COSMOS&0.3725&s&9.53& & \\
   &150.14990234	&2.2571671&COSMOS&0.3715&s&8.81&87 &5.2:1\\
   \hline
\enddata
\tablecomments{(1) System ID; (2) Right ascension; (3) Declination; (4) Field name; (5) Redshift; (6) Source of redshift (s-spectroscopic, p-photometric). For photometric redshifts 95\% confidence limits are reported from \cite{2016ApJS..225...27M}; (7) Stellar mass in log(M$_*$/\(\textup{M}_\odot\)); (8) Projected separation in kpc between the most massive dwarf and its neighbor(s); (9) Mass ratios between the most massive dwarf and its neighbor(s).}
\end{deluxetable*}

\startlongtable
\begin{deluxetable*}{ccccccccc}
\tablenum{2}
\tablecaption{Summary of spectroscopically confirmed dwarf groups\label{tab2}}
\tablewidth{0pt}
    \tablehead{\colhead{ID} & \colhead{RA}  & \colhead{Dec} & \colhead{Field}&\colhead{$z$} & \colhead{$z$ source} &\colhead{log M$_*$}&\colhead{r$_{\textup{sep}}$}&\colhead{M$_{1*}$/M$_{2*}$} \\
    \colhead{} & \colhead{(deg)}  & \colhead{(deg)} & \colhead{}&\colhead{} & \colhead{} &\colhead{(\(\textup{M}_\odot\))}&\colhead{(kpc)}&\colhead{}  \\ \hline
    \colhead{(1)} & \colhead{(2)}  & \colhead{(3)} & \colhead{(4)}&\colhead{(5)} & \colhead{(6)} &\colhead{(7)}&\colhead{(8)}&\colhead{(9)}
}
\startdata
id52   &189.09373474	&62.16689682& GOODS-N&0.7504&s&9.58& & \\
   &189.09140015	&62.16662598&GOODS-N&0.7504&s&9.11&30&2.9:1 \\
   &189.09063721&	62.16683197&GOODS-N&0.7504&s&8.76&39&6.6:1 \\ \hline
id53   &215.00132751&	52.9636879& AEGIS&0.7444&s&9.56&&  \\
   &215.00022888	&52.96072006&AEGIS&0.7439&s&9.06&89&3.2:1 \\
   &214.99899292	&52.96024323&AEGIS&0.7437&s&8.98&90 &3.8:1\\ \hline
id54   &189.32987976	&62.25835037& GOODS-N&0.8419&s&9.39&&  \\
   &189.33276367	&62.258979&GOODS-N&0.8421&s&9.25&32 &1.4:1\\
   &189.33218384	&62.2587204&GOODS-N&0.8423&s&8.90&41&3.1:1 \\ \hline
id55   &189.24627686&	62.31162643& GOODS-N&1.0110&s&9.59&&  \\
   &189.24359131	&62.31165314&GOODS-N&1.0111&s&9.56&37&1.1:1 \\
   &189.24624634	&62.31002808&GOODS-N&1.0120&s&9.11&47&3:1 \\ \hline
id56   &189.12358093&	62.20001221& GOODS-N&1.0153&s&9.39& & \\
   &189.1244812	&62.2002449&GOODS-N&1.0160&s&9.00&14&2.4:1 \\
   &189.12409973	&62.20019531&GOODS-N&1.0160&s&8.80&9&3.1:1 \\ \hline
id57   &214.98876953	&52.85660934& AEGIS&0.3340&s&9.14&&  \\
   &214.98529053	&52.85574341&AEGIS&0.3342&s&9.00&39&1.4:1 \\ 
   &214.98249817	&52.85441971&AEGIS&0.3335&s&8.43&76&5.1:1\\
   \hline
id58   &189.31768799	&62.22966385& GOODS-N&0.4475&s&9.73& & \\
   &189.31654358&	62.23021317&GOODS-N&0.4466&s&8.58&16&14.1:1 \\
   &189.31764221	&62.23020554&GOODS-N&0.4465&s&8.52&11&16.2:1 \\
   \hline
id59   &189.40246582&	62.33492661& GOODS-N&0.5857&s&9.40&&  \\
   &189.40385437	&62.33591843&GOODS-N&0.5856&s&9.11&28&1.9:1 \\
   &189.39904785&	62.33481979&GOODS-N&0.5857&s&9.07&38&2.1:1 \\ 
   &189.40063477&	62.33504105&GOODS-N&0.5656$^{+0.1204}_{-0.1366}$&p&7.98&21&26.3:1\\
   \hline
\enddata
\tablecomments{(1) System ID; (2) Right ascension; (3) Declination; (4) Field name; (5) Redshift; (6) Source of redshift (s-spectroscopic, p-photometric). For photometric redshifts 95\% confidence limits are reported from \cite{2016ApJS..225...27M}; (7) Stellar mass in log(M$_*$/\(\textup{M}_\odot\)); (8) Projected separation in kpc between the most massive dwarf and its neighbor(s); (9) Mass ratios between the most massive dwarf and its neighbor(s).}
\end{deluxetable*}

\startlongtable
\begin{deluxetable*}{ccccccccc}
\tablenum{3}
\tablecaption{Summary of candidates for dwarf-dwarf pairs and groups\label{tab3}}
\tablewidth{0pt}
    \tablehead{\colhead{ID} & \colhead{RA}  & \colhead{Dec} & \colhead{Field}&\colhead{$z$} & \colhead{$z$ source} &\colhead{log M$_*$}&\colhead{r$_{\textup{sep}}$}&\colhead{M$_{1*}$/M$_{2*}$} \\
    \colhead{} & \colhead{(deg)}  & \colhead{(deg)} & \colhead{}&\colhead{} & \colhead{} &\colhead{(\(\textup{M}_\odot\))}&\colhead{(kpc)}&\colhead{}  \\ \hline
    \colhead{(1)} & \colhead{(2)}  & \colhead{(3)} & \colhead{(4)}&\colhead{(5)} & \colhead{(6)} &\colhead{(7)}&\colhead{(8)}&\colhead{(9)}
}
\startdata
id60   &189.22039795	&62.28197098& GOODS-N&0.7614&s&8.77&&  \\
   &189.22059631	&62.28224182&GOODS-N&0.7610$^{+0.1010}_{-0.0730}$&p&8.54&8&1.7:1 \\
 \hline
id61   &150.11117554	&2.25387931& COSMOS&0.7767&s&9.61&&  \\
   &150.11143494&	2.25390387&COSMOS&0.7771$^{+0.0024}_{-0.0030}$&p&8.84&7&5.9:1 \\
   \hline
id62   &189.17056274	&62.27451324& GOODS-N&0.7802&s&9.71&&  \\
   &189.16990662&	62.27450562&GOODS-N&0.7816$^{+0.0034}_{-0.0053}$&p&9.41&8&2:1 \\
    \hline
id63   &189.17514038	&62.22260284
& GOODS-N&0.8460&s&9.15&&  \\
   &189.17657471	&62.22203827&GOODS-N&0.8487$^{+0.0040}_{-0.0027}$&p&9.03&24&1.3:1 \\
 \hline
id64   &189.32949829	&62.35944748& GOODS-N&0.9458&s&9.43&  &\\
   &189.32531738	&62.35821152&GOODS-N&0.9455$^{+0.0046}_{-0.0043}$&p&9.24&66&1.5:1 \\ \hline
id65   &53.13801193	&--27.69859123& GOODS-S&0.9523$^{+0.0019}_{-0.0038}$&p&9.46& & \\
   &53.13689041&	--27.69883156&GOODS-S&0.9510&s&9.18&29&1.9:1 \\
   \hline
id66   &214.8289032&	52.83329773& AEGIS&0.9582$^{+0.0024}_{-0.0030}$&p&9.34&&  \\
   &214.82546997	&52.83427429&AEGIS&0.9600&s&9.27&66&1.2:1 \\
   \hline
id67   &53.17535019	&--27.85260391& GOODS-S&0.9616&s&9.39& & \\
   &53.17354202	&--27.85242081&GOODS-S&0.9630$^{+0.0029}_{-0.0126}$&p&9.15&47&1.7:1 \\
   \hline
id68   &53.07076645&	--27.85066795& GOODS-S&0.9900&s&9.59&&  \\
   &53.0722084	&--27.84975815&GOODS-S&0.9917$^{+0.0182}_{-0.0099}$&p&9.12&45&2.9:1\\
   \hline
id69   &214.92391968	&52.82198715
& AEGIS&0.8390&s&9.03& & \\
   &214.92401123	&52.82240295&AEGIS&0.8368$^{+0.0043}_{-0.0042}$&p&8.68&12&2.2:1\\
   \hline
id70   &189.27220154&	62.17921829
& GOODS-N&1.0230&s&9.46& & \\
   &189.27285767&	62.17891693&GOODS-N&1.0262$^{+0.1128}_{-0.1592}$&p&8.28&13&15.1:1\\
   \hline
id71   &189.14463806	&62.16677856
& GOODS-N&1.2111$^{+0.0082}_{-0.0175}$&p&9.18&&  \\
   &189.14497375	&62.16666412&GOODS-N&1.2174&s&9.14&8&1.1:1\\
   \hline
id72   &53.08846283	&--27.76522827
& GOODS-S&1.2240&s&9.22& & \\
   &53.08879471	&--27.76517105&GOODS-S&1.2233$^{+0.0016}_{-0.0038}$&p&8.87&9&2.2:1\\
   \hline
id73   &53.16807938&	--27.72351074
& GOODS-S&1.2434&s&9.59&&  \\
   &53.16835785	&--27.72373772&GOODS-S&1.2453$^{+0.0857}_{-0.1013}$&p&9.02&10&3.7:1\\
   \hline
id74   &53.11321259	&--27.83273125
& GOODS-S&1.2930&s&9.64& & \\
   &53.11421585&	--27.83185387&GOODS-S&1.2945$^{+0.0022}_{-0.0033}$&p&9.10&38&3.5:1\\
   \hline
id75   &189.13853455&	62.23644638
& GOODS-N&1.3448&s&9.52& & \\
   &189.13601685	&62.23694611&GOODS-N&1.3486$^{+0.1054}_{-0.1206}$&p&9.02&39&3.2:1\\
   \hline
id76   &189.10520935	&62.18302917
& GOODS-N&1.3500&s&9.55& & \\
   &189.10534668	&62.18320084&GOODS-N&1.3446$^{+0.0914}_{-0.1036}$&p&9.04&6&3.2:1\\
   \hline
id77   &53.10016632	&--27.84272003
& GOODS-S&1.4130&s&9.71& & \\
   &53.09990311	&--27.84256554&GOODS-S&1.4177$^{+0.2593}_{-0.1167}$&p&8.81&9&7.9:1\\
   \hline
id78   &189.16906738	&62.23484421
& GOODS-N&1.4805$^{+0.2225}_{-0.1405}$&p&9.28& & \\
   &189.16880798	&62.23432541&GOODS-N&1.4856&s&9.13&17&1.4:1\\
   \hline
   id79   &189.30671692	&62.33421707& GOODS-N&0.2777&s&8.61&&  \\
   &189.30674744&	62.33452988
&GOODS-N&0.2770$^{+0.0054}_{-0.0056}$&p&7.89&5&5.2:1 \\
 \hline
id80   &189.29902649	&62.25413895& GOODS-N&0.2990&s&9.56&&  \\
   &189.30000305	&62.25340652&GOODS-N&0.3028$^{+0.0328}_{-0.0477}$&p&8.19&14 &23.4:1\\
   \hline
id81   &53.16155624	&--27.79225731& GOODS-S&0.4580&s&9.11& & \\
   &53.16196442	&--27.79255295&GOODS-S&0.4528$^{+0.0315}_{-0.0106}$&p&8.79&10&2.1:1 \\
    \hline
id82   &53.22396469&	--27.81417465
& GOODS-S&0.1483&s&8.16& & \\
   &53.22394943	&--27.81487083&GOODS-S&0.1434$^{+0.0389}_{-0.0188}$&p&7.63&6&3.4:1 \\
 \hline
id83   &150.08700562	&2.3090229& COSMOS&0.6875&s&9.19& & \\
   &150.08738708	&2.3089335&COSMOS&0.6846$^{+0.0049}_{-0.0080}$&p&8.87&10&2.1:1 \\ \hline
id84   &53.09656143	&--27.92811012& GOODS-S&0.7334&s&9.50& & \\
   &53.09663391	&--27.92639351&GOODS-S&0.7350$^{+0.0077}_{-0.0061}$&p&8.79&46&5.1:1 \\
   \hline
id85   &214.75852966	&52.75759506& AEGIS&0.8456&s&9.63& & \\
   &214.75782776	&52.75798035&AEGIS&0.8404$^{+0.0024}_{-0.0034}$&p&9.05&16&3.8:1 \\
   \hline
id86   &53.15167999	&--27.77537727& GOODS-S&1.0391$^{+0.0061}_{-0.0189}$&p&9.71& & \\
   &53.15179062	&--27.77550697&GOODS-S&1.0470&s&9.09&5&4.2:1 \\
   \hline
id87   &150.06022644	&2.28915334& COSMOS&0.6041&s&9.38& & \\
   &150.06088257	&2.28896928&COSMOS&0.5968$^{+0.0390}_{-0.0203}$&p&9.03&17&2.2:1\\
   \hline
id88   &214.88000488&	52.87176514& AEGIS&0.6913&s&9.13&&  \\
   &214.8795929	&52.87166214&AEGIS&0.6930$^{+0.1076}_{-0.0192}$&p&8.77&7&2.3:1\\
   \hline
id89   &214.96824646	&52.91167831& AEGIS&0.7684&s&9.31&  &\\   &214.96798706&52.9119873&AEGIS&0.7674$^{+0.0578}_{-0.0096}$&p&8.91&9&2.5:1\\
   \hline
   id90   &189.3994751	&62.24066925
& GOODS-N&1.1403&s&9.70&&  \\   
&189.39570618	&62.23989487
&GOODS-N&1.1436$^{+0.0431}_{-0.0926}$&p&9.44&58&1.8:1\\
   \hline
id91   &53.14494324	&--27.84020996
& GOODS-S&0.9778&s&9.42&&  \\
   &53.1451683	&--27.84038353&GOODS-S&0.9818$^{+0.0032}_{-0.0029}$&p&9.42&8&1:1\\
   &53.14426041&	--27.84013176
&GOODS-S&0.9807$^{+0.0553}_{-0.0717}$&p&8.89&18&3.4:1\\
   \hline
id92   &53.18664551	&--27.83599281
& GOODS-S&1.2960&s&9.59&&  \\
   &53.18647766	&--27.83611488
&GOODS-S&1.2960&s&9.31&6&1.9:1\\
   &53.18729401&	--27.83481789
&GOODS-S&1.3002$^{+0.0022}_{-0.0034}$&p&9.52&40&1.2;1\\
   \hline
id93   &53.19279099	&--27.91669655
& GOODS-S&1.4330$^{+0.0730}_{-0.0700}$&p&9.52& & \\
   &53.19335938	&--27.91612625
&GOODS-S&1.4242$^{+0.0021}_{-0.0032}$&p&9.00&23&3.3:1\\
   &53.19394302&	--27.91622162
&GOODS-S&1.4250&s&9.12&34&2.5:1\\
   \hline
\enddata
\tablecomments{(1) System ID; (2) Right ascension; (3) Declination; (4) Field name; (5) Redshift; (6) Source of redshift (s-spectroscopic, p-photometric). For photometric redshifts 95\% confidence limits are reported from \cite{2016ApJS..225...27M}; (7) Stellar mass in log(M$_*$/\(\textup{M}_\odot\)); (8) Projected separation in kpc between the most massive dwarf and its neighbor(s); (9) Mass ratios between the most massive dwarf and its neighbor(s).}
\end{deluxetable*}
\begin{deluxetable*}{ccccccccc}
\tablenum{4}
\tablecaption{Summary of X-ray detected AGN\label{tab4}}
\tablewidth{0pt}
    \tablehead{\colhead{ID} & \colhead{RA}  & \colhead{Dec} & \colhead{log L$_X$}&\colhead{log M$_*$} & \colhead{log SFR} &\colhead{log L$_{XRB}$}&\colhead{log L$_{hot}$}&\colhead{log L$_{AGN}$} \\
    \colhead{} & \colhead{(deg)}  & \colhead{(deg)} & \colhead{(erg s$^{-1}$)}&\colhead{(\(M_\odot\))} & \colhead{(\(M_\odot\) yr$^{-1}$)} &\colhead{(erg s$^{-1}$)}&\colhead{(erg s$^{-1}$)}&\colhead{(erg s$^{-1}$)}  \\ \hline
    \colhead{(1)} & \colhead{(2)}  & \colhead{(3)} & \colhead{(4)}&\colhead{(5)} & \colhead{(6)} &\colhead{(7)}&\colhead{(8)}&\colhead{(9)}
}
\startdata
id3   & 53.0798	&--27.7462   & 40.18&8.69&0.18& 39.40$\pm$0.07 &39.09$\pm$0.01&40.06$\pm$0.02\\ \hline
id10   & 53.0658	&--27.7750   & 41.54&9.33&1.17& 40.38$\pm$0.06 &39.98$\pm$0.01&41.50$\pm$0.02\\ \hline
id34   & 53.0444	&--27.7875   & 41.21&8.93&-0.10& 39.13$\pm$0.06 &38.7$\pm$0.01&41.20$\pm$0.01\\ \hline
id43   & 53.1259	&--27.7514   &43.56&9.55&1.22&  40.43$\pm$0.06&40.04$\pm$0.01&43.56$\pm$0.00\\ \hline
id86   & 53.1518	&--27.7755   & 41.74&9.09&1.35&  40.56$\pm$0.06&40.17$\pm$0.01&41.70$\pm$0.00\\ \hline
id91   & 53.1450	&--27.8403   & 41.96&9.42&1.01&  40.22$\pm$0.06&39.83$\pm$0.01&41.86$\pm$0.00\\ \hline 
\enddata
\tablecomments{(1) Corresponding system ID; (2) X-ray source right ascension; (3) X-ray source declination; (4) K-corrected X-ray luminosity calculated adopting the redshift of the host galaxy as distance and using the aperture-corrected net energy flux from the point spread function 90\% enclosed-count fraction aperture in the rest-frame 0.5-10 keV in units erg s$^{-1}$ cm$^{-2}$. For id3, id10, id34, id43, and id91 reported X-ray flux is derived from the longest observation; for id87 reported X-ray flux is derived by averaging over all contributing observations; (5) Stellar mass of the host galaxy (6) Star formation rate of the host galaxy adopted from \cite{2016ApJS..225...27M}(7) X-ray binary contribution to the observed X-ray flux calculated using \cite{Lehmer_2010}; (8) Hot interstellar medium contribution to the observed X-ray flux calculated using \cite{10.1111/j.1365-2966.2012.21831.x} (9) Corrected AGN X-ray luminosity.}
\end{deluxetable*}

%



\clearpage
\bibliography{sample631}{}
\bibliographystyle{aasjournal}



\end{document}